\documentclass[twocolumn,aps]{revtex4}
\usepackage{amsmath}
\usepackage{graphics}

\begin{document}
  
\title{Influence of solvent quality on effective pair potentials
  between polymers in solution}
  
\author{V. Krakoviack\footnote{Present address: Laboratoire de Chimie,
\'Ecole Normale Sup\'e\-rieure de Lyon, 46, All\'ee d'Italie, 69364 Lyon
Cedex 07, France.}, J.-P. Hansen, and A. A. Louis}
  
\affiliation{Department of Chemistry, University of Cambridge,
  Lensfield Road, Cambridge CB2 1EW, United Kingdom}
 
\date{\today}

\begin{abstract}
Solutions of interacting linear polymers are mapped onto a system of
``soft'' spherical particles interacting via an effective pair
potential. This coarse-graining reduces the individual monomer-level
description to a problem involving only the centers of mass (CM) of
the polymer coils. The effective pair potentials are derived by
inverting the CM pair distribution function, generated in Monte Carlo
simulations, using the hypernetted chain (HNC) closure.  The method,
previously devised for the self-avoiding walk model of polymers in
good solvent, is extended to the case of polymers in solvents of
variable quality by adding a finite nearest-neighbor monomer-monomer
attraction to the previous model and varying the temperature.  The
resulting effective pair potential is found to depend strongly on
temperature and polymer concentration.  At low concentration the
effective interaction becomes increasingly attractive as the
temperature decreases, eventually violating thermodynamic stability
criteria. However, as polymer concentration is increased at fixed
temperature, the effective interaction reverts to mostly repulsive
behavior. These issues help illustrate some fundamental difficulties
encountered when coarse-graining complex systems via effective pair
potentials.
\end{abstract}

\maketitle

\section{Introduction}
While the computer simulation of single, isolated polymer chains,
either on or off lattice, using a variety of conformation sampling
algorithms, is nowadays relatively routine, for up to $L\simeq10^6$
monomers or segments \cite{review}, it is computationally much more
demanding to simulate polymer solutions or melts, involving large
numbers of interacting polymer chains. Indeed, if $N$ is the number of
such chains and $L$ the number of segments per polymer, then the total
number of interacting particles, $NL$, can become very large,
particularly so if $L$ is sufficiently large for the scaling regime to
be reached \cite{degennes}. Under these conditions it is tempting to
seek a coarse-graining procedure to reduce the full segment-level
description to a model involving only the center of mass (CM) or the
central monomer of each chain, thus reducing the initial $NL$-body
problem to a $N$-body problem. This is formally achieved by tracing
out the individual monomer degrees of freedom, i.e. by averaging over
polymer conformations for fixed positions of the CM's or central
monomers of interacting polymer coils, taking into account the
appropriate Boltzmann weights. This idea goes back to Flory and
Krigbaum \cite{flokri50jcp}. They predicted that the effective
repulsive interaction between the CM's of linear self-avoiding walk
(SAW) polymers should diverge with molecular weight at full overlap,
i.e. when the CM's of the two coils coincide. It was first realized by
Grosberg \textit{et al.} \cite{grokhakho82mcrc} that in fact the pair
potential between CM's remains finite in the scaling limit
$L\to\infty$, and of the order of a few $k_BT$, reflecting the purely
entropic origin of the effective interaction. The effective CM pair
potential between two isolated SAW polymer coils was explicitly
evaluated by Monte Carlo (MC) simulations of on and off-lattice models
\cite{olalanpel80mm,dauhal94mm,bollouhanmei01jcp}, and by
renormalization group (RG) calculations \cite{kruschbau89jpf}. These
studies show that the effective pair potential is purely repulsive,
with an overlap (zero separation) value of about $2k_BT$ and a range
of the order of the radius of gyration $R_g$. The zero concentration
pair potential is reasonably well represented by a single Gaussian of
width $R_g$.

More recently the effective pair potential was determined at
\emph{finite} polymer concentration by a combination of monomer-level
MC simulations of lattice SAW polymers, and an inversion technique
based on integral equations for pair distribution functions in simple
liquids \cite{simpleliquids}. The resulting pair potentials depend now
moderately on polymer concentration
\cite{bollouhanmei01jcp,bollou02mm}, but they remain essentially
repulsive and of range $R_g$. They have been put to good use to
reproduce the interfacial tension of semi-dilute polymer solutions
near hard walls or colloidal spheres \cite{loubolmeihan02jcp1}, and to
determine the depletion interaction between colloidal particles
induced by interacting (rather than ideal) polymer coils
\cite{loubolmeihan02jcp2} and the resulting, depletion-induced phase
diagram of colloidal dispersions \cite{bollouhan}.

The present paper reports an extension of the above inversion strategy
to the case of dilute and semi-dilute solutions of interacting linear
polymers in solvents of variable quality, spanning the range between
good solvent conditions, modeled by the SAW, and poor solvent
conditions, where the coils contract to avoid contact with the
solvent. This generalization is achieved by adding a finite attractive
interaction between nearest-neighbor monomers of the same or different
chains, while maintaining infinite repulsion between overlapping
monomers. The finite nearest-neighbor attraction introduces an energy
scale, and hence a temperature dependence of the effective pair
potential. Good solvent conditions correspond to the infinite
temperature limit, leading back to the SAW model considered earlier,
while increasingly poor solvent conditions are mimicked by enhancing
the attraction between monomers, or equivalently, decreasing the
temperature.

The dependence of the properties of a single polymer coil, like its
radius of gyration $R_g$, on temperature, and in particular around and
below the $\theta$ temperature where the coil to globule transition
takes place, has been studied by extensive MC simulations of very long
chains ($L$ up to $10^6$) by Grassberger and collaborators
\cite{graheg95jcp}. Results for the temperature dependence of the
effective pair potential for an off-lattice model of two polymer coils
(i.e. in the infinite dilution limit) were reported by Dautenhahn and
Hall \cite{dauhal94mm}. These authors met with increasing MC sampling
difficulties as the temperature is decreased towards the $\theta$
point. The present paper improves on their results by considering a
lattice model, and using an overlapping distribution method
\cite{ben76jcp,understanding} to sample rare conformations which
become increasingly important as the temperature decreases. The main
objective of this paper is to consider, for the first time, the
effective pair potential between the CM's of polymer coils in poor
solvent at \textit{finite} polymer concentration, i.e. to investigate
both the temperature and concentration dependence of the CM pair
potential, in particular in the vicinity of the $\theta$ point. The
polymer model studied in this work is defined in Section \ref{model}.
The zero concentration limit is first considered in Section
\ref{zerodens}, while finite polymer concentrations are examined next
in Section \ref{finitedens}. Some considerations on the use of
effective interactions in relation to thermodynamic stability are
presented in Section \ref{stability}, and conclusions are drawn in
Section \ref{conclusion}.

\section{Model and basic phenomenology}\label{model}

For all the calculations in the present work, a lattice model defined
on the simple cubic lattice (with a coordination number of six) was
used \cite{lattice}. A polymer chain of length $L$ is represented as a
connected sequence of $L$ lattice sites. The monomers, defined as
occupied sites, interact with each other via excluded volume
repulsion, preventing two of them from occupying the same lattice
position, and lattice nearest-neighbor attraction $\varepsilon<0$
between non-bonded pairs of monomers. For convenience, the temperature
$T$ will be expressed in units of $-\varepsilon/k_B$, or equivalently
we will set $\varepsilon=-1$ and $k_B=1$. With this convention, the
usual $\beta$ appearing in the Boltzmann factor is simply the inverse
of the temperature, i.e. $\beta=1/T$.  In the following, we shall call
``contact'' any nearest-neighbor pair of non-bonded monomers. We use
``intramolecular'' if both monomers belong the same chain, and
``intermolecular'' otherwise. When a system consisting of $N$ chains
of $L$ segments is considered on a portion of cubic lattice of $M$
sites, the monomer packing fraction is equal to the fraction of
lattice sites occupied by polymer segments, $c=NL/M$, while the
concentration of polymer chains is $\rho=c/L=N/M$.

\begin{figure}
\includegraphics{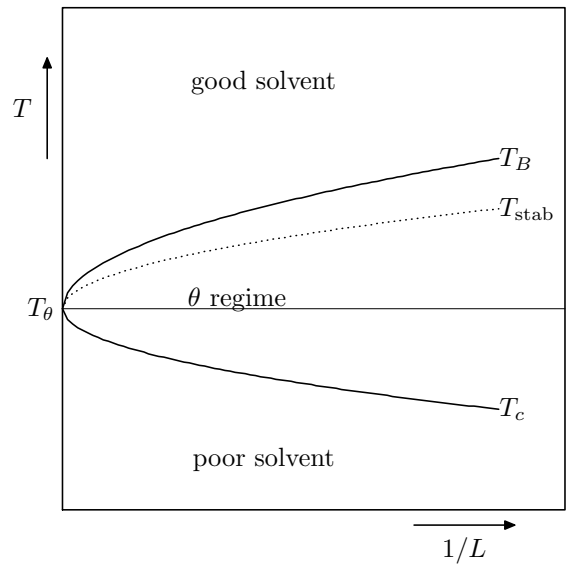}
\caption{\label{sketch} Schematic temperature-inverse length diagram
showing the three regimes of polymer phenomenology and the behavior of
the temperatures $T_\theta$, $T_B$, $T_{\text{stab}}$, and $T_c$
(defined in the text) related to the characterization of the $\theta$
regime.}
\end{figure}

This simple model has been extensively studied in order to investigate
the properties of isolated polymer chains and of polymer solutions. In
the phenomenology of these systems, three different domains, sketched
on Fig.~\ref{sketch}, are usually defined, corresponding to different
behaviors of an isolated chain.

In the good solvent regime, corresponding in the present model to high
temperatures, the properties of the system are essentially determined
by the entropic effects originating in connectivity and excluded
volume interactions. An isolated chain takes swollen  coil
configurations with a radius of gyration $R_g$ known to scale as
$L^{\nu}$, where $\nu \simeq 0.588$ is the Flory exponent in three
dimensions \cite{flory}. This regime has been extensively studied in
the case of the SAW to which the present model reduces in the infinite
temperature limit \cite{limadsok95jsp}.

In the bad or poor solvent regime, corresponding here to low
temperatures, entropic effects are overwhelmed by the attractive
monomer-monomer interactions and a chain molecule collapses into a
compact globule with a radius of gyration scaling like $L^{1/3}$.

Between these two extremes lies the $\theta$ regime, where energetic
and entropic effects compensate each other. In this regime, in the
limit of infinite length, a chain has the scaling properties of an
ideal chain, for instance $R_g\propto L^{1/2}$. The characterization
of this domain has been the focus of many studies using the present
lattice model \cite{bru84mm,meilim90jcp,szlotopan92jcp,yandep00jcp},
with in particular extensive work by Grassberger and coworkers
\cite{graheg95jcp,gra97pre,fragra97jcp}.

Of key importance is the so-called $\theta$ temperature, $T_\theta$,
at which the tricritical coil-to-globule transition takes place for an
isolated chain in the $L\to\infty$ limit. It corresponds to a real
thermodynamic singularity and its best estimate is presently
\cite{gra97pre}
\begin{equation*}
T_\theta=3.717\pm0.003\text{, i.e. } \beta_\theta=0.2690\pm0.0002. 
\end{equation*}
Obviously, simulations can only deal with chains of finite length, for
which the tricritical singularity is rounded off by significant finite
size corrections. $T_\theta$ is thus not directly accessible and it is
common practice, in simulations as well as in experiments, to obtain
it by extrapolation of related, length-dependent characteristic
temperatures.

The most widely used of such temperatures is Boyle's temperature,
$T_B(L)$, defined as the temperature at which the osmotic second
virial coefficient of chains of length $L$ vanishes. From its
knowledge, one can compute the $\theta$ temperature from the limit
\begin{equation}
\lim_{L\to\infty} T_{B}(L) = T_\theta.
\end{equation}
This route has been followed in particular in Ref.~\cite{graheg95jcp},
where it was found that $T_B(L)$ is always greater than $T_\theta$ and
decreases with increasing $L$.

An alternative possibility is to use the demixing critical temperature
of the polymer solution. This marks the limit between a high
temperature regime where the chains in solution form a homogeneous
fluid, and a low temperature domain where the chains start to
coagulate and the fluid demixes into polymer-rich and polymer-poor
phases. In the lattice model used here, where the solvent is taken
into account implicitly through the introduction of constant
attractive monomer-monomer interactions and variations of the
temperature, it is simply the critical temperature $T_c(L)$ of the
liquid-gas transition of the system. As for $T_B(L)$, one has
\begin{equation}
\lim_{L\to\infty} T_{c}(L) = T_\theta,
\end{equation}
and it has been found in Refs.~\cite{yandep00jcp} and
\cite{fragra97jcp} that $T_{c}(L)$ is always smaller than $T_\theta$
and increases with increasing $L$.

We anticipate on the following sections and introduce here a third
characteristic temperature called the ``stability'' temperature and
denoted $T_{\text{stab}}(L)$. It is associated with a fundamental
breakdown in the statistical-mechanical treatment of the
coarse-grained effective pair interaction $v_2(r)$ between the CM's of
two polymer chains at low temperature. Singularities of a similar
nature have been encountered in various models for soft matter
systems. For instance, Baxter's sticky sphere model \cite{bax68jcp}
displays anomalous clustering towards a closed-packed crystalline
phase as a result of breakdown of thermodynamic stability
\cite{ste91jsp}. In the case of polymers, the Domb-Joyce model
\cite{domjoy72jpc} shows self-trapping behavior, i.e.  a chain remains
of finite extent in the infinite length limit, when a negative energy
is attributed to self-crossings \cite{duxquesti84jpa}.

The relation between $T_{\text{stab}}(L)$ and the $\theta$ point is
unclear. For any length $L$, $T_{\text{stab}}(L)$ will be shown to be
smaller than $T_B(L)$, the equality being only achieved when the pair
interaction is zero. For the specific length $L=100$ studied in detail
in this paper, it is found to fall between $T_B(L)$ and
$T_\theta$. Because it corresponds to a singularity in the
statistical-mechanical treatment of the system, it is tempting to
assume, as it was done in Ref.~\cite{loubolhan00pre}, that
\begin{equation}
\lim_{L\to\infty} T_{\text{stab}}(L) = T_\theta,
\end{equation}
and Fig.~\ref{sketch} has been drawn under this assumption.
Interestingly, this would imply that, in the infinite length limit,
since $T_B(L\to\infty)=T_{\text{stab}}(L\to\infty)=T_\theta$, the
effective interaction between two isolated polymer chains at the
$\theta$ point vanishes identically for all distances, a behavior
which is trivially found in the case of the Domb-Joyce model
\cite{domjoy72jpc}, when a positive constant energy penalty is counted
for each polymer crossing (note that in this model, the temperature scale is
reversed since the SAW is obtained at zero temperature and ideal chain
behavior is found at infinite temperature).

\section{Zero density limit}\label{zerodens}
\subsection{Simulation methodology}

In the zero density limit, the effective interaction potential
$v_2(r)$ between two polymer chains is equal to the difference between
the free energy of the two chains with their centers of mass
constrained to stay at a fixed distance $r$, and the free energy of
the same chains infinitely far apart. It can be expressed as follows.

Consider two polymer chains labeled $A$ and $B$. When these chains
have conformations $\Gamma_A$ and $\Gamma_B$ respectively, with the
vector $\mathbf{r}_{AB}$ joining their centers of mass, the energy of
the pair is
\begin{multline}
H(\mathbf{r}_{AB};\Gamma_A,\Gamma_B)=H_{\text{intra}}(\Gamma_A)\\
+H_{\text{intra}}(\Gamma_B)+
H_{\text{inter}}(\mathbf{r}_{AB};\Gamma_A,\Gamma_B),
\end{multline}
where the intra- and intermolecular parts (including the hard-core
interactions) have been separated. Introducing the intermolecular
Boltzmann weight
\begin{equation}\label{boltzweight}
W(|\mathbf{r}_{AB}|;\Gamma_A,\Gamma_B)=\exp\left[-\beta\,H_{\text{inter}}
(\mathbf{r}_{AB};\Gamma_A,\Gamma_B)\right],
\end{equation}
the effective pair potential is given by \cite{ensemble}
\begin{equation}\label{meanforce}
\beta\,v_2(r)\equiv-\ln \langle W(|\mathbf{r}_{AB}|=r; 
\Gamma_A,\Gamma_B)\rangle,
\end{equation}
where the brackets denote an average over the probability distribution
of two isolated chains, which is the square of the probability
distribution $\mathcal{P}$ of a single chain, i.e.
\begin{multline}
\langle W(|\mathbf{r}_{AB}|=r;\Gamma_A,\Gamma_B)\rangle=\\
\sum_{\Gamma_A,\Gamma_B} \mathcal{P}(\Gamma_A)\mathcal{P}(\Gamma_B)
W(|\mathbf{r}_{AB}|=r;\Gamma_A,\Gamma_B) 
\end{multline}
with
\begin{equation}
\mathcal{P}(\Gamma)=\exp\left[-\beta\,H_{\text{intra}}(\Gamma)\right]\left/
\sum_\Gamma \exp\left[-\beta\,H_{\text{intra}}(\Gamma)\right]\right..
\end{equation}

This result provides us with a direct means to compute the effective
pair potential between two chains by Monte Carlo simulations. We
sample configurations of two independent chains using the pivot
algorithm \cite{madsok88jsp} and standard Metropolis acceptance rules.
The latter ensure that the chain conformations are generated according
to the probability distribution $\mathcal{P}$. After every 1000 pivot
moves for each chain, we calculate the intermolecular Boltzmann weight
\eqref{boltzweight} as a function of the CM distance, by moving the
polymers towards each other, while checking for overlap and counting
intermolecular contacts. Eventually, $\beta\,v_2(r)$ is obtained by
performing the unweighted average of
$W(|\mathbf{r}_{AB}|=r;\Gamma_A,\Gamma_B)$ on the sample considered:
\begin{equation}
\beta\,v_2(r)=-\ln\frac{\displaystyle\sum_{i=1}^{\mathcal{N}(r)} 
W(|\mathbf{r}_{AB}|=r;\Gamma_A^i,\Gamma_B^i)}{\mathcal{N}(r)},
\end{equation}
where $\mathcal{N}(r)$ is the number of two chain configurations with
CM's at distance $r$ sampled during the simulation.

Using this algorithm, we have been able to compute with good
statistical accuracy effective pair potentials for various chain
lengths ($L=100$, $200$ and $500$), provided $\beta\le0.2$.

Unfortunately, this simple, direct method turns out to be inadequate
at lower temperatures, where the results for the effective potential
at short distances are found to fluctuate strongly with the sample
considered, with no significant reduction of the corresponding
statistical uncertainties when the size of this sample is increased.
The origin of this problem, already seen in the analogous off-lattice
calculation of Dautenhahn and Hall \cite{dauhal94mm}, has been
carefully analyzed by Grassberger and Hegger in their calculation of
the second virial coefficient of the same lattice model as the one
studied here \cite{graheg95jcp}. At low temperature, the polymer
chains start to collapse significantly and are thus rather compact.
This means that at short distances, most two-chain configurations
obtained with the present direct algorithm will overlap and the
corresponding intermolecular Boltzmann weights are then identically
zero. But, in the rare cases where the two chains do not overlap, many
contacts will usually be formed, leading to very low negative
intermolecular energies and thus huge intermolecular Boltzmann weights
contributing to the pair potential. The need to average such unevenly
distributed numbers of very different amplitudes gives rise to the
large observed fluctuations and renders the direct approach useless.

Hence more elaborate algorithms must be used for low temperatures. In
order to choose these new tools, it is helpful to recognize the strong
similarity between the previous scheme and Widom's particle insertion
method for the computation of the chemical potential \cite{wid63jcp}.
Both methods indeed involve the averaging of a Boltzmann weight over
the equilibrium distribution of some unperturbed system, and in fact
the observed breakdown of the direct calculation of the effective pair
potential at low temperature parallels that of Widom's method at high
fluid densities. In the latter case, an efficient approach to solve
the problem has been devised by Shing and Gubbins \cite{shigub82mp},
which belongs to the general class of overlapping distribution methods
first introduced by Bennett to compute free-energy differences
\cite{ben76jcp,understanding}. We have thus implemented such a method
which is described in detail in the Appendix.

Using this histogram method, we have been able to extend our
calculation of the effective pair potentials between chains of length
$L=100$ to lower temperatures, i.e. up to $\beta=0.3$. However, the
method breaks down when applied to longer chains, for mainly two
reasons.  Firstly, the acceptance ratio of the elementary Monte Carlo
move becomes extremely small, leading to serious ergodicity problems
in the simulations. Secondly, except for small separations
corresponding to nearly complete overlap of the chains, the two
computed histograms do not overlap at all, rendering the estimation of
the required free energy difference quite problematic. The origin of
this problem can easily be understood on a qualitative level. In the
original system, the chains attract each other strongly and lower
their intermolecular energy at low temperature by elongating towards
each other along the axis joining their CM's: there is thus a sizeable
volume where the two chains overlap with the consequent formation of
contacts in large numbers \cite{grokuz92mm}. On the contrary, in the
reference system, only unfavorable excluded volume interactions exist
between the two partly collapsed chains which are thus compressed
along the axis joining their CM's to limit their overlap
\cite{dauhal94mm}: contacts are then formed in small numbers only at
the surface separating the two segregated chains. This significant
separation in the number of contacts does not occur (or at least is
less pronounced for the lengths studied here) at complete overlap
because the system has then to keep the spherical symmetry of the
isolated chains, thus leaving no possibility for the adverse changes
of the chain shapes seen in the cylindrical symmetry.

Although it would be possible to use alternative methods (see
e.g. \cite{graheg95jcp,yandep00jcp}) to sample even longer chains at
low temperature, we use mainly $L=100$ chains in this paper, since we
do not expect the qualitative features we focus on here to depend
strongly on chain length \cite{semidilute}. Of course, quantitative
features will vary with $L$, and the detailed scaling behavior can be
quite complex \cite{gra97pre,fragra97jcp}.

\subsection{Results}

\begin{figure*}
\includegraphics{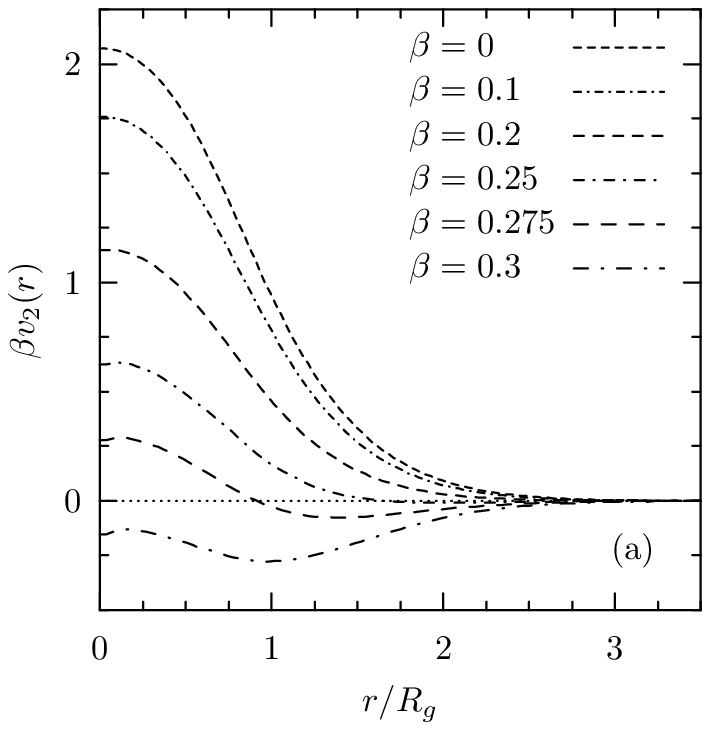}
\hspace{0.5cm}
\includegraphics{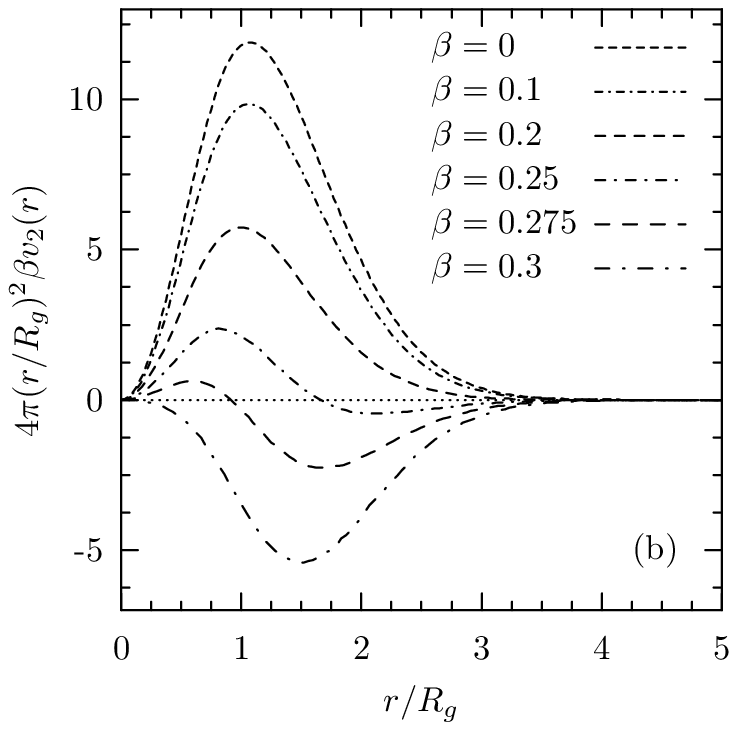}
\caption{\label{direct} Zero density effective pair potentials at
different inverse temperatures $\beta$ for polymers of length $L=100$
on the simple cubic lattice. At each temperature, the $x$-axis is
scaled with the corresponding $R_g(\beta)$, which is shown in
Fig.~\ref{rgzero} }
\end{figure*}

\begin{figure}
\includegraphics{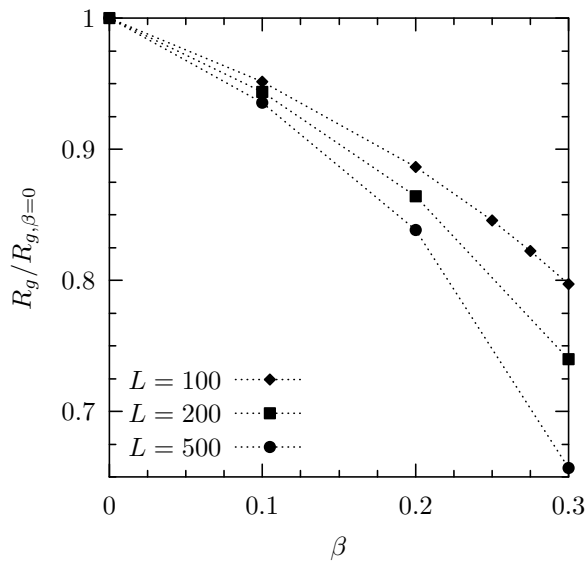}
\caption{\label{rgzero} Radii of gyration of isolated polymer chains
of different lengths $L$ on the simple cubic lattice as functions of
the inverse temperature $\beta$.  The data are scaled with the
infinite temperature values, $R_g(L=100)=6.44$, $R_g(L=200)=9.76$, and
$R_g(L=500)=16.84$.}
\end{figure}

The effective pair interactions $\beta v_2(r)$ between two polymers of
length $L=100$ are plotted, for various solvent qualities, in
Fig.~\ref{direct}. The product of the potential by $4\pi r^2$ is shown
as well, because the integral of this quantity plays an important role
for the thermodynamics of the system
\cite{loubolhan00pre,lanlikwatlowjpcm00}, and to emphasize the
features of the potential at large distances. In this figure and in
all the following figures representing effective pair potentials, all
distances have been scaled at each temperature by the radius of
gyration of the isolated chain at this temperature, which is plotted
in Fig.~\ref{rgzero} for completeness.

One can distinguish two temperature domains. In
Ref.~\cite{bollouhanmei01jcp}, where the infinite temperature limit of
the model was studied, $\beta v_2(r)$ was found to be purely repulsive
with a roughly Gaussian shape centered on $r=0$. Here we find that it
remains so, provided the temperature is high enough, i.e.
$\beta\lesssim 0.2$. The main effect of lowering the temperature is a
reduction of the overall amplitude of the potential and a slight
decrease in its range in terms of the normalized distance.

At lower temperatures, $\beta>0.2$, $\beta v_2(r)$ begins to exhibit
qualitative changes: while the potential retains a repulsive
Gaussian-like component at short distance, a negative, attractive tail
appears at large distance. As the temperature decreases further, the
amplitude of the repulsive core decreases, following the trend of the
previous high temperature domain, whereas the attractive tail becomes
more and more important, ultimately dominating the whole picture, as
seen for $\beta=0.3$, where $\beta v_2(r)$ is everywhere negative and
only a modest repulsion shows up for $r/R_g<1$.

A second, less prominent feature appears at low temperature
($\beta\gtrsim0.2$) as well. At $r=0$, the effective pair potential
displays a small minimum, which becomes deeper when the temperature
decreases. This means that full overlap is locally stable and that one
has to overcome a (modest) free-energy barrier to separate two chains
in this configuration. For chains of length $L=100$, it is difficult
to know if this feature is a generic property of the effective pair
potential or just a lattice artifact due to the shortness of the
chains (the width of the minimum is indeed of the order of the lattice
spacing). However, despite our limited ability to investigate longer
chains at low temperature, we have seen evidence of a similar minimum,
but of larger width, for chains of length $L=500$ at $\beta=0.225$,
suggesting that this is a genuine physical effect.

One can compute various interesting scalar quantities from the
knowledge of the effective pair potential. According to
Sec.~\ref{model}, the second osmotic virial coefficient, given by
\begin{equation}\label{eqvirialcoeff}
B_2(\beta)=\frac{1}{2}\int_0^\infty\left[1-\exp(-\beta\,v_2(r))\right]
4\pi r^2dr,
\end{equation}
is of particular importance. In addition, we introduce the
``stability'' integral,
\begin{equation}\label{stabilityint}
I_2(\beta)=\frac{1}{2}\int_0^\infty\beta\,v_2(r)4\pi r^2dr,
\end{equation}
which results from the linearization of the exponential in
Eq.~\eqref{eqvirialcoeff}. For systems interacting through
\emph{density-independent} potentials, the sign of $I_2$ gives a
necessary condition for the existence of the thermodynamic limit and
the stability of the system against coalescence
\cite{ruelle,gallavotti}, as will be further developed in
Sec.~\ref{stability}. Accordingly, as discussed by anticipation in
Sec.~\ref{model}, a ``stability'' temperature $T_{\text{stab}}(L)$ can
be defined, at which $I_2(\beta)$ vanishes for chains of length $L$,
and below which the necessary condition for stability against
coalescence is violated. The ordering $T_{\text{stab}}(L) \leq
T_{B}(L)$ discussed above is then an immediate consequence of the fact
that $x\ge 1-\exp(-x)$.

\begin{figure}
\includegraphics{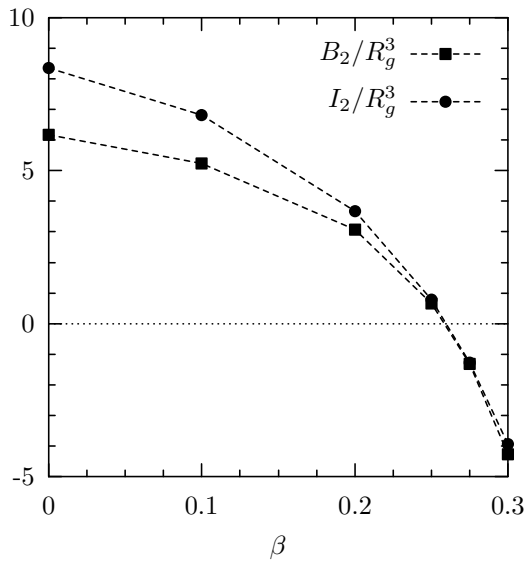}
\caption{\label{vir} Reduced second osmotic virial coefficient
($B_2/R_g^3$) and reduced stability integral at zero density
($I_2/R_g^3$) as functions of the inverse temperature $\beta$ for
polymers of length $L=100$ on the simple cubic lattice.}
\end{figure}

Both quantities, normalized with the cube of the radius of gyration to
obtain dimensionless quantities, are plotted in Fig.~\ref{vir}. As was
implicitly assumed above, they are decreasing functions of $\beta$
and one finds, for the chains of length $L=100$ studied here, that
$T_B\gtrsim T_{\text{stab}}\simeq 3.8$.

\begin{figure}
\includegraphics{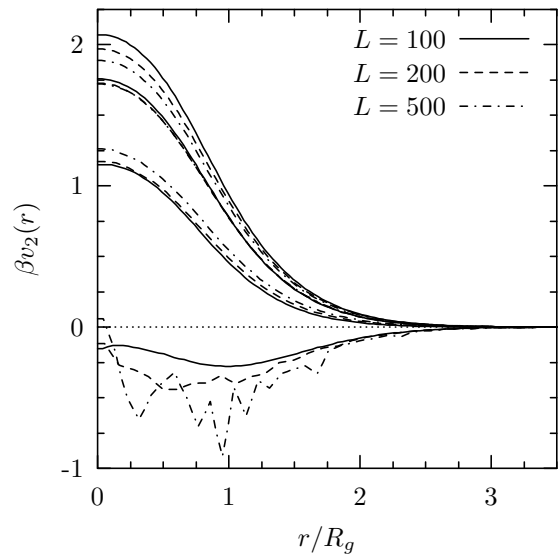}
\caption{\label{length} Zero density effective pair potentials at
different inverse temperatures $\beta$ (from top to bottom, for a
given length, $\beta=0$, $0.1$, $0.2$, and $0.3$) for polymers of
various lengths $L$ on the simple cubic lattice. At each temperature,
the $x$-axis is scaled with the corresponding $R_g(\beta)$.}
\end{figure}

Finally, we address the question of the dependence of the zero density
effective pair potential on the chain length. In
Ref.~\cite{bollouhanmei01jcp}, the infinite temperature case has been
studied in detail over a wide range of polymer lengths ($L=100$ to
$8000$). Such an extensive study was not attempted here and we only
considered two rather short lengths ($L=200$ and $500$) in addition to
the previous data for $L=100$. All these data, plotted in
Fig.~\ref{length} with the one previously described, have been
computed with the direct method, leading to strong statistical
uncertainties at the lowest temperature corresponding to $\beta=0.3$.

To interpret these potentials one should keep in mind that $R_g$
scales differently with temperature for the different lengths, as
illustrated in Fig.~\ref{rgzero}.  Also, $T_B$ and $T_c$ are different
for each length.  Following Refs.~\cite{graheg95jcp} and
\cite{fragra97jcp}, in which these temperatures have been computed
with great care, for $L=100$: $T_c \simeq 3.10$; $T_B \simeq 3.86$;
for $L=200$: $T_c \simeq 3.33$; $T_B \simeq 3.81$; for $L=500$: $T_c
\simeq 3.41$; $T_B \simeq 3.77$.

A few qualitative observations can be made, at least for $\beta\leq
0.2$, where the statistics are good enough. With the normalization
chosen for the $x$-axis, no intersection between potentials
corresponding to different values of $L$ at the same temperature is
found. In Ref.~\cite{bollouhanmei01jcp}, this feature was already
found for $\beta=0$, combined with the fact that a larger $L$ leads to
a less repulsive potential. We recover this result here and the same
qualitative behavior is found for $\beta=0.1$, but the potentials for
different values of $L$ are closer. This behavior changes for
$\beta=0.2$, where we now see that the larger $L$, the more repulsive
the potential. Globally, we thus find that in the high-temperature
domain, the decrease in the amplitude of the repulsive effective
potential is slower for large values of $L$.

It is difficult to draw conclusions from the data at $\beta=0.3$, but
it looks like the larger $L$, the deeper the attractive potential.
Indeed, for large CM distances, at which the statistical inaccuracies
in the simulation results are expected to be modest, the potential
seems to be more negative when $L$ is large; the same holds at
complete overlap: trying to estimate $\beta v_2(0)$ for $L=500$ with
the overlapping distribution method, we find $\beta v_2(0)\simeq-1.1$,
to be compared to $\beta v_2(0)\simeq-0.15$ for $L=100$.

All these results, if applied to Eq.~\eqref{eqvirialcoeff}, are
consistent with the findings of Grassberger and Hegger for the
variation of the reduced second osmotic virial coefficient,
$B_2/R_g^3$, with temperature for various lengths (see Fig.~16 in
Ref.~\cite{graheg95jcp}). Indeed, they found that the larger $L$, the
flatter this quantity is in the high temperature regime, the more
abrupt the downward bend of the curve when approaching $T_B(L)$ and
the faster the divergence towards $-\infty$ in the poor solvent
regime.

\section{Finite densities}\label{finitedens}
\subsection{Methodology}

Having derived the effective potential between two isolated polymers,
we now turn to polymers at finite density. For this, we follow the
route proposed in previous work \cite{bollouhanmei01jcp}, in which an
effective pair potential was constructed to exactly reproduce the
two-body CM correlations of the full underlying many-body system. In
fact, it can be proven for a wide variety of systems that for any
given pair distribution function $g(r)$ at given inverse temperature
$\beta$ and density $\rho$, there exists a corresponding unique
two-body pair potential $v(r)$ which reproduces $g(r)$ irrespective of
the underlying many-body interactions in the system
\cite{hen74pla}. Of course, $g(r)$ depends on density and temperature
and contains contributions not only from the bare pair potential
$v_2(r)$, but also from the three- and more-body terms. As a
consequence, the effective pair interaction will also be state
dependent and a new effective potential, hereafter denoted
$v(r;\rho,\beta)$, must be calculated for each density and
temperature.  However this inversion approach says nothing about a
possible volume term $v_1(\rho,\beta)$, in the coarse-grained total
potential energy, which contributes to the equation of state, but not
\emph{directly} to the pair-correlations \cite{gralow98pre,Liko01}. Of
course the volume terms may still contribute \emph{indirectly}, for
example when they induce phase transitions.

The inversion procedure, using $g(r)$ to extract $v(r)$, is well known
and has been studied extensively in the field of simple fluids
\cite{rea86pma,zerhan86jcp}. We invert $g(r)$ using the
hypernetted chain (HNC) closure,
\begin{equation}
\label{eq:hnc}
g(r)=\exp(-\beta v(r)+g(r)-c(r)-1),
\end{equation}
of the Ornstein-Zernike equation \cite{simpleliquids}. While the
simple HNC inversion procedure would be inadequate for dense fluids of
hard core particles, where more sophisticated closures or iterative
procedures are required \cite{rea86pma,zerhan86jcp}, we are able to
demonstrate the consistency of the HNC inversion in the present case.

To compute the necessary structural information we have performed
canonical Monte Carlo simulations of polymer solutions. We have
studied chains of length $L=100$ in a cubic box of size $M=100^3$ with
periodic boundary conditions, varying the number of polymers from
$N=400$ to $N=3200$ ($c=0.04$ to $0.32$). Four temperatures have been
considered corresponding to $\beta=0$, $0.1$, $0.2$, and $0.3$. Note
that the lowest temperature, $T=3.33$, is slightly larger than the critical
temperature of the system, $T_c(L=100)\simeq 3.1$ according to
Ref.~\cite{fragra97jcp}, thereby avoiding concerns with the possible
two-phase behavior of the system. To sample the configuration space of
the system we have used standard techniques of polymer simulations: pivot
moves \cite{madsok88jsp}, translation moves, and, for high densities
where the previous moves become inefficient, configurational bias
Monte Carlo (CBMC) moves, in which an extremity or part of the
interior of a chain are regrown \cite{understanding}.

In the course of the simulations, the CM of each polymer was tracked
in order to construct the CM radial pair distribution function
$g(r;\beta,\rho)$. The latter is only known up to a cutoff radius
$r_c$, which corresponds to half the size of the simulation box. For
the inversion, we need $g(r)$ for all $r$, so we employ the following
iterative scheme to extend $g(r)$. As an initial step, we set $g(r)=1$
for $r>r_{c}$ and calculate the corresponding $v(r)$ by inversion. We
then set $v(r)=0$ for $r>r_{c}$ and determine the corresponding $g(r)$
for $0<r<\infty$ by a regular HNC calculation, using a simple
iterative procedure. The $g(r)$ for $r<r_{c}$ is then replaced by the
measured $g(r)$, and the new $v(r)$ is calculated.  This is again set
to zero for $r>r_{c}$, and the process is repeated until
convergence. In fact, because of the finite box-size, the inversion
process is underdetermined, and our ansatz that $v(r)=0$ for $r > r_c$
is needed to find a unique solution \cite{bollou02mm}. This is not
unreasonable since we do not expect the interactions between the
polymer coils to be significant beyond distances a few times the
radius of gyration. However, to make sure that this is actually the
case, we found that relatively large simulation boxes were needed,
with a lattice size of up to $10$-$15R_g$. This is particularly
important at high density, where the inverted potential becomes longer
ranged and more sensitive to small changes in the radial distribution
function $g(r)$. In all our inversions, we checked explicitly that
$v(r)$ becomes effectively zero before $r=r_c$, confirming our initial
ansatz.

\subsection{Results}

\begin{figure*}
\includegraphics{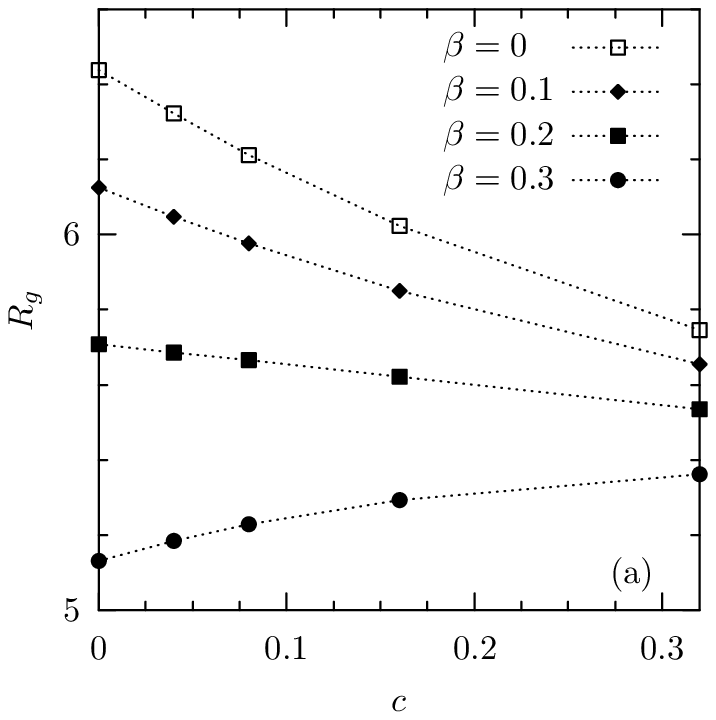}
\hspace{0.5cm}
\includegraphics{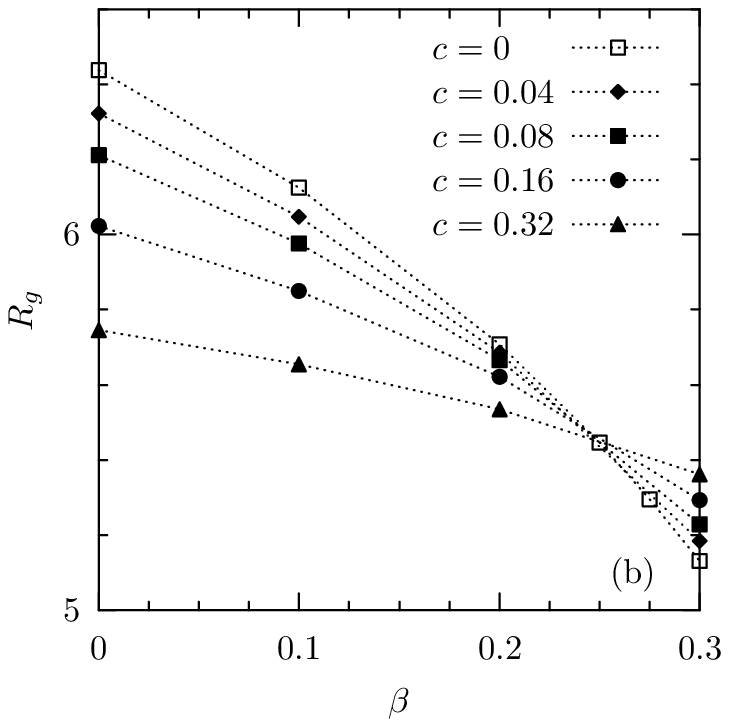}
\caption{\label{rg} Density- and temperature-dependence of
  the radius of gyration of polymers of length $L=100$ on the simple
  cubic lattice. (a) Radii of gyration as functions of the monomer
  density $c$ for various inverse temperatures $\beta$. (b) Radii of
  gyration as functions of the inverse temperature $\beta$ for various
  monomer densities $c$.}
\end{figure*}

Before presenting the results for the effective pair potentials, we
first discuss briefly the behavior of the radius of gyration with
density and temperature, which is an important measure of the physical
properties of the polymer.  The corresponding simulation data are
plotted in Fig.~\ref{rg}. As found in similar previous work
\cite{milpaubin93jcp}, for a given temperature, $R_g$ decreases when
the density of polymer increases if this temperature falls into the
good solvent regime, whereas $R_g$ increases if this temperature is
located below the $\theta$ regime (Fig.~\ref{rg}a). When $R_g$ is
plotted as a function of temperature for various densities, this
results in the existence of a regime, located in the $\theta$ region,
where $R_g$ is nearly independent of $c$, as can be seen in
Fig~\ref{rg}b, where all the curves are seen to converge around
$\beta\simeq 0.25$.

\begin{figure*}
\includegraphics{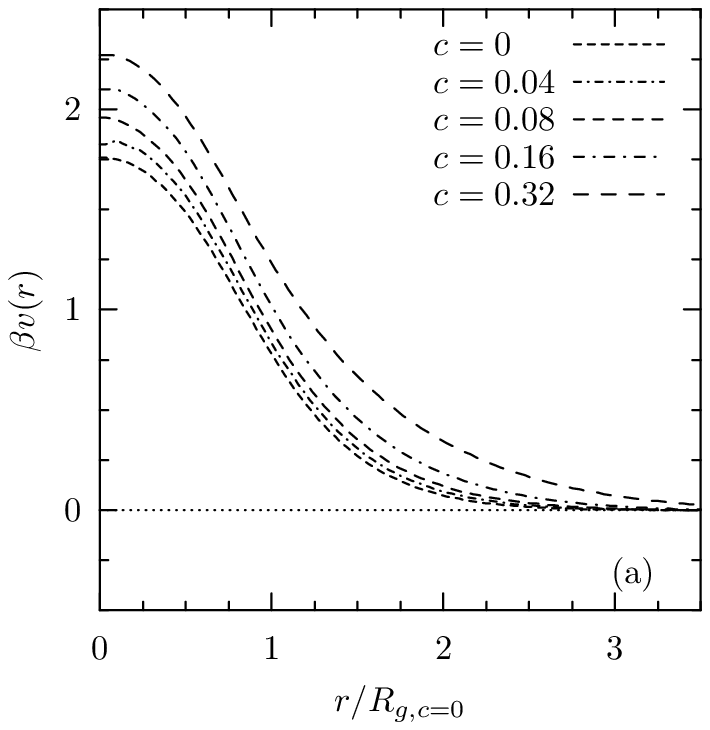}
\hspace{0.5cm}
\includegraphics{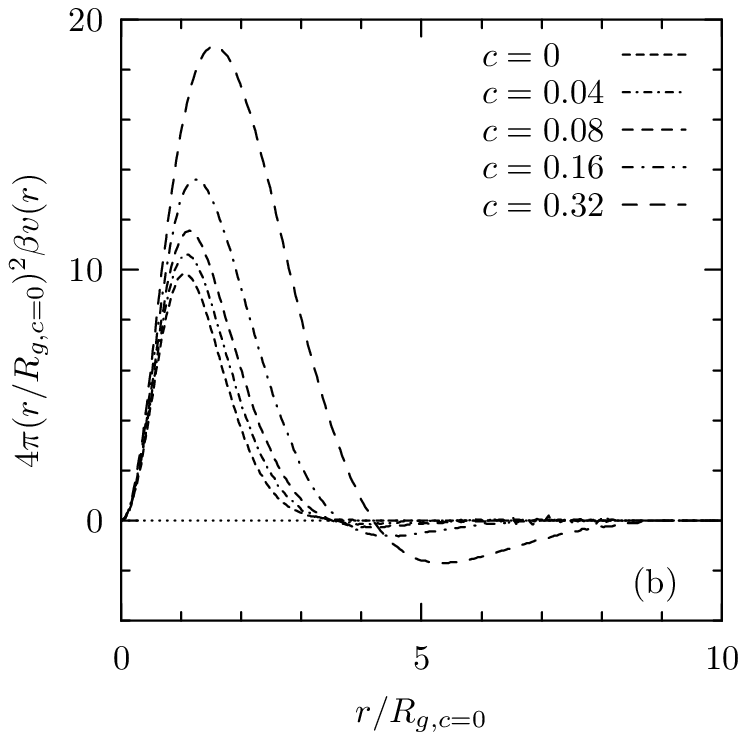}
\caption{\label{dens1} Effective pair potentials for polymers of
length $L=100$ on the simple cubic lattice at the inverse temperature
$\beta=0.1$ and at various monomer densities $c$. The $x$-axis is
scaled with the radius of gyration corresponding to $c=0$ and
$\beta=0.1$.}
\end{figure*}

\begin{figure*}
\includegraphics{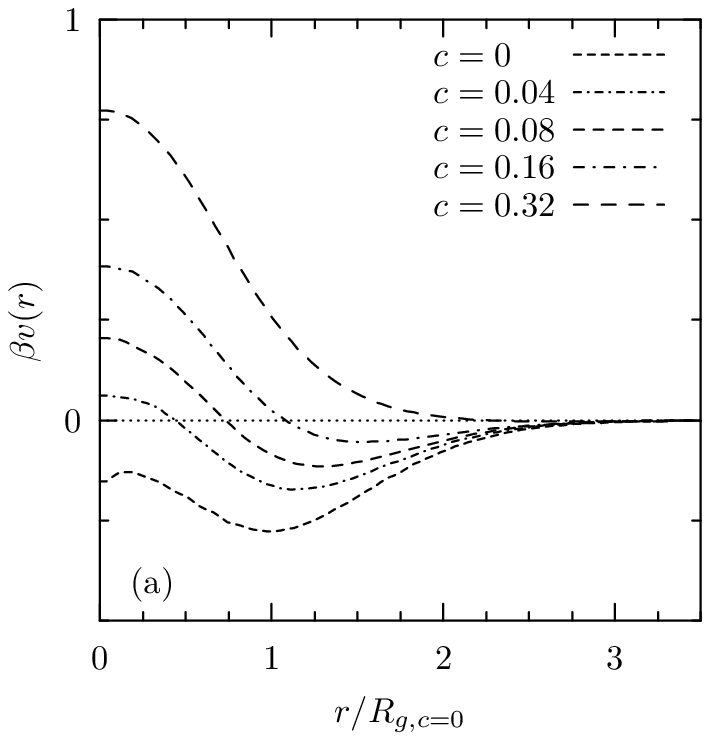}
\hspace{0.5cm}
\includegraphics{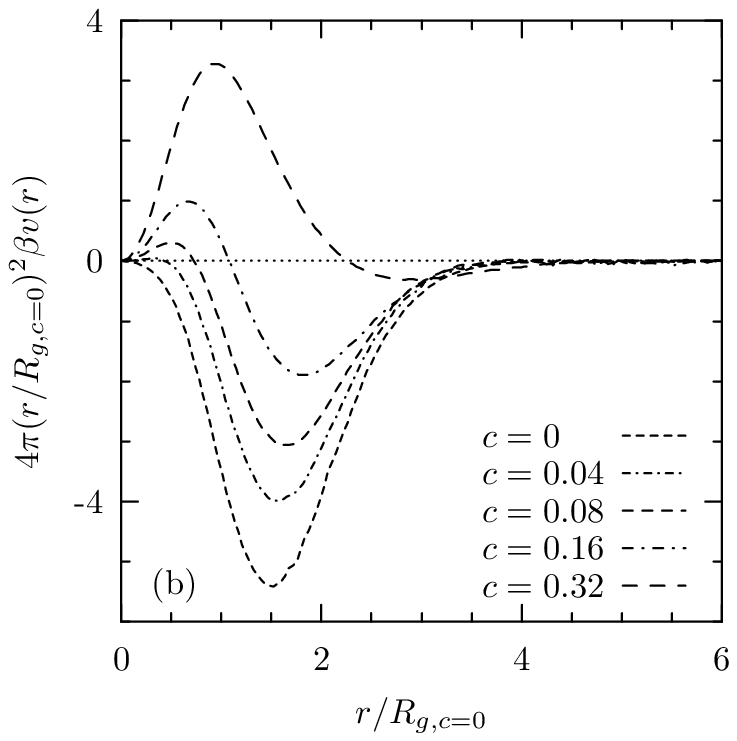}
\caption{\label{dens3} Effective pair potentials for polymers of
length $L=100$ on the simple cubic lattice at the inverse temperature
$\beta=0.3$ and at various monomer densities $c$. The $x$-axis is
scaled with the radius of gyration corresponding to $c=0$ and
$\beta=0.3$.}
\end{figure*}

The effective temperature- and density-dependent pair potentials
are plotted in Figs.~\ref{dens1} and \ref{dens3}  for
two representative temperatures, corresponding to $\beta=0.1$ and
$\beta=0.3$ respectively. From these data, it is clear that the two
temperature domains identified in the zero density study of
Sec.~\ref{zerodens} display obvious distinctive features in terms of
the density dependence of the effective pair potentials at fixed
temperature.

When $\beta=0.1$ (Fig.~\ref{dens1}), i.e.~in the high temperature
regime, a moderate density dependence is found. Starting from the
purely repulsive, nearly Gaussian shape of the potential at zero
density and increasing the density, one finds a slight increase in the
amplitude and in the range of the repulsive potential for small
separations of the polymer CM's and the progressive appearance of a
weak negative attractive tail at large distance. This behavior is
similar to the one found in the infinite temperature limit studied in
Ref.~\cite{bollouhanmei01jcp}, with the only minor difference that we
do not see a decrease of the amplitude of the potential at short
distances for the highest concentrations.

The situation is completely different when $\beta=0.3$
(Fig.~\ref{dens3}). Here the density dependence of the effective pair
potential is very important and leads to significant qualitative
changes in the shape of the potential: whereas at low density, $v(r)$
is negative and essentially attractive, repulsion between centers of
mass progressively builds up when the density increases, so that an
almost purely repulsive, Gaussian shaped $v(r)$ is eventually observed
at the largest density.  For the good solvent case a clear link has
been found between the density dependence and the strength of the 3-
and higher body interactions \cite{bollouhan01pre,lou02jpcm}.
Presuming that the same link can be made here, this would imply that
the many-body interactions are relatively more important  at lower
temperatures.

The virial coefficient has no clear interpretation for potentials
obtained at finite global densities, whereas the ``stability''
integral does still define a lower limit to the existence of a
thermodynamic limit of the coarse-grained system.  As can be directly
seen from Fig.~\ref{dens3}b, at all densities, except the largest
($c=0.32$), the stability integral of the computed potentials is
negative, leading to serious consistency issues to be discussed in the
next section.

\section{Considerations of thermodynamic stability}
\label{stability}

The potentials computed in this work raise a number of interesting
conceptual issues with regard to the use of effective potentials in
coarse-grained descriptions of material properties. The first question
concerns \emph{thermodynamic stability} \cite{ruelle}: do the
calculated pair potentials generate a valid thermodynamic limit ?
Secondly, there is a question of \emph{representability}
\cite{lou02jpcm}: for a given state point, how well can the properties
of the underlying polymer system be coherently represented by a single
coarse-grained effective pair potential ?

\subsection{Thermodynamic stability of effective potentials}

We briefly repeat the criteria for the existence of a thermodynamic
limit described in Ruelle's classic book \cite{ruelle} and valid for
state independent interactions. Consider a system of $N$ particles in
a volume $V$. If the total interaction energy $V_N$, which can be
built from pair and higher order terms, satisfies, for all $N > 0$ and
for all configurations $\{\mathbf{r}_i\}$ in the configuration space
$R^N$, the inequality
\begin{equation}\label{eqA1}
V_N(\mathbf{r}_1,\ldots,\mathbf{r}_N) \geq - B N
\end{equation}
with $B \geq 0$, then, according to definition $3.2.1.$ in
Ref.~\cite{ruelle}, the system is \emph{stable}: the grand partition
function converges, and there is a well-defined thermodynamic limit.
Potentials that do not satisfy this criterion are termed
\emph{catastrophic} by Ruelle. Specializing to pair potentials, i.e.
\begin{equation}\label{eqA2}
V_N^{(2)}(\mathbf{r}_1,\ldots,\mathbf{r}_N) = \sum_{1 \leq i < j \leq N}
v_2(|\mathbf{r}_i - \mathbf{r}_j|),
\end{equation}
condition~\eqref{eqA1} leads to the following necessary (but not
sufficient) condition for the existence of a thermodynamic limit:
\begin{equation}\label{eqA3}
\hat{v}(0)= \int v_2(r) 4\pi r^2dr > 0,
\end{equation}
which, following Eq.~\eqref{stabilityint}, can be rewritten $I_2>0$,
hence the name ``stability'' integral for $I_2$. If Eq.~\eqref{eqA3}
is not satisfied, then condition~\eqref{eqA1} can be violated for
configurations of a homogeneous fluid. As $N \rightarrow \infty$, the
free energy grows super-extensively -- the system has no thermodynamic
limit -- and particles coalesce to form a dense
cluster. Condition~\eqref{eqA3} is necessary, but not sufficient, for
the existence of a thermodynamic limit, since one can also construct
potentials for which $\hat{v}(0) > 0$, but where the system in a
microscopically inhomogeneous (typically, crystalline) state is
unstable to coalescence. See Refs.~\cite{ruelle} and
\cite{loubolhan00pre} for explicit examples.

Effective potentials with hard cores, such as those used to describe
simple atomic and molecular materials, can be easily shown to satisfy
the criterion~\eqref{eqA1}. In contrast, the potentials describing the
effective interactions between the CM's of polymers studied in the
present work do not have a hard core, leading to the possibility that
two or more ``effective'' particles occupy the same position in space.
The existence of a thermodynamic limit, where the free energy per
particle is bounded, is therefore much more subtle.

Although the potentials calculated in this work are state dependent
and technically only relevant at the density for which they have been
obtained, it is nevertheless interesting to consider what would happen
should one of them be used, independently of the density, to describe
a system of $N>2$ particles in a volume $V$. To this end, we first
consider the zero density potentials shown in Fig.~\ref{direct}.

When the underlying SAW polymer system has no nearest-neighbor monomer
attractions, the effective pair potentials were shown to be positive
\cite{loubolhan00pre,bollouhanmei01jcp}, so that the criterion
\eqref{eqA1} is obeyed. However, as Fig.~\ref{direct} demonstrates,
the introduction of nearest-neighbor attractions leads to effective
potentials which are no longer positive definite. As the temperature
is lowered, the potential grows more and more attractive, until
finally it violates the Ruelle criterion and becomes
catastrophic. This can be diagnosed in Fig.~\ref{vir}, where $I_2$
becomes negative at low temperature and the necessary condition
\eqref{eqA3} is thus violated.  In principle, the stability limit of
the system should be traced by looking for a state point at which the
potential leads to coalescence into microscopically inhomogeneous
states, a rather difficult task in general. Here we will use a
simpler, approximate criterion, namely Eq.~\eqref{eqA3}, which is a
necessary, but not sufficient condition for stability.  We will make
the heuristic assumption that the true stability limit is not far
removed from the simpler one we employ \cite{more}.  For a given
potential then, Eq.~\eqref{eqA3} allows us to define the ``stability''
temperature $T_{\text{stab}}$ below which the effective pair potential
becomes catastrophic.  This explains our nomenclature in
Secs.~\ref{model} and \ref{zerodens}.

The potentials derived at zero density violate Eq.~\eqref{eqA3} below
a temperature $T_{\text{stab}} \approx 3.8$; the potentials are
expected to become unstable to inhomogeneous coalescence at some
temperature above that.  If one were to use potentials derived for $T
\leq T_{\text{stab}}$ at a finite density, the system would be
catastrophic.

This naturally leads to the next question: what about the potentials
we derived at finite density ?  There, as shown in Fig.~\ref{dens3},
the pair potential can also violate Eq.~\eqref{eqA3}, even though the
underlying polymer system is stable, both to coalescence and phase
separation, since $T> T_c $ for our $L=100$ SAW lattice chains.  For
large enough densities $c$, the potentials again satisfy
Eq.~\eqref{eqA3}.  

Serious difficulties are thus emerging with the present
coarse-graining procedure: in certain portions of the
temperature-density plane, we replace a complex, but perfectly
well-behaved, polymer system by a simple soft colloid fluid with
pathological thermodynamics!  This also raises a certain number of
formal concerns.  The first one is about the unicity of the mapping
between $g(r)$ and $v(r)$ invoked in Sec.~\ref{finitedens} and
demonstrated in Ref.~\cite{hen74pla}. The theorems rely on
well-defined statistical ensembles and their derivations are no longer
valid when a catastrophic potential is the outcome of their
application.  Similarly, the lack of a well defined ensemble also
leads to questions about the validity of the liquid state theory,
including the Ornstein-Zernike equation, etc.

However, in spite of these formal difficulties, one could take a
purely pragmatic attitude: Since it is numerically possible to extract
effective pair potentials using the proposed procedure, why not just
ignore all the previous concerns and see if these potentials can be of
any practical use ? To investigate this point, MC simulations of a soft
colloid system interacting through the catastrophic effective pair
potential obtained for polymers at $\beta=0.3$ and $c=0.04$ were
performed, revealing a very interesting behavior.

\begin{figure}
\includegraphics{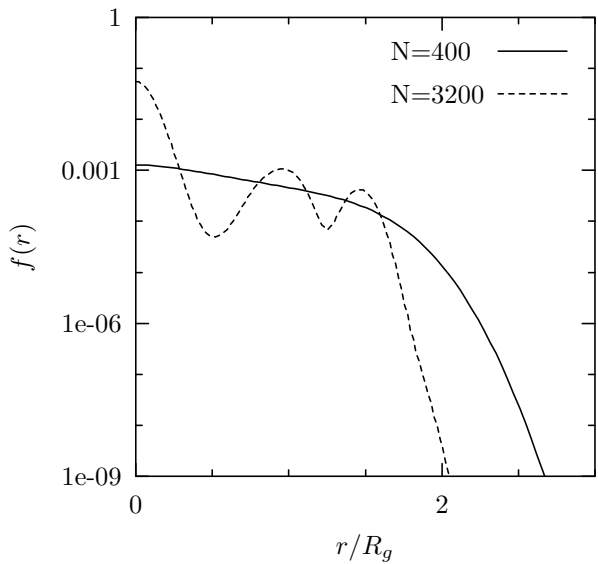}
\caption{\label{cluster} Pair distribution functions of two collapsed
soft colloid systems, with $N=400$ and $N=3200$ particles
respectively. In both cases, the space integral of $f(r)$ has been
normalized to 1 to allow a more significant comparison. The
interaction potential is the effective pair potential obtained for a
polymer system at $\beta=0.3$ and $c=0.04$.}
\end{figure}

We indeed find that the behavior of the soft colloid fluid depends
strongly on the configuration from which the simulation is initiated.
When an initial configuration is set up with all soft particles
gathered in a single dense cluster, the system coalesces: the
particles stay together indefinitely (after a few million MC steps per
particle, not a single one has been seen to escape the cluster), and
the potential energy per particle $e$ appears unbounded, increasing in
absolute value with increasing $N$; more precisely, we find $e\simeq
-24.6$ for $N=400$ and $e\simeq -202$ for $N=3200$, which would give
$e(N)\simeq -0.062N$. This is exactly the type of behavior expected
from a system interacting through a catastrophic potential.  The
structure of the clusters also shows distinctive features, as can be
seen in Fig.~\ref{cluster}, where pair distribution functions $f(r)$
are plotted for the previous cluster sizes. For $N=3200$, $f(r)$
displays three peaks, located at $r/R_g=0$, $r/R_g\simeq1$, and
$r/R_g\simeq1.5$. Quite evidently, the first two peaks originate in
the formation of groups of superimposed particles separated by the
distance corresponding to the interaction potential minimum. This is
indeed a very efficient way for the system to lower its energy, since
the energy cost for overlapping particles is modest compared to the
stabilization of pairs of particles at distance $r\simeq R_g$. As for
the third peak, it is very likely it has to be associated with the
position of the second nearest neighbors. In comparison, for $N=400$,
$f(r)$ is rather featureless and longer-ranged. This change in shape
of $f(r)$ is again easily understood from the clustering
mechanism. For small $N$, smaller groups of superimposed particles can
be formed, creating less deep energy wells at distance $r\simeq R_g$;
the resulting clusters are thus more diffuse and less structured than
for large $N$.

But, if the initial conditions correspond to a homogeneous
distribution of particles in the simulation box, the fluid is found to
remain homogeneous over the entire time of our simulations (up to
$45.10^6$ MC steps per particle for 400 particles), suggesting that
the homogeneous fluid is metastable with respect to catastrophic
coalescence. In this case, the potential energy per particle $e$ is,
as expected, independent of the size of the system,
i.e. $e\simeq-0.176$ for systems containing $400$ or $3200$ particles.

Of course, the original polymer system, which is perfectly stable,
does not show such a dependence of its behavior on the initial
conditions. This can be easily checked by preparing an initial
configuration consisting of a dense cluster of polymer chains: the
cluster ``evaporates'' very rapidly, and the density becomes uniform
again within a few thousand MC steps per chain only.

\begin{figure}
\includegraphics{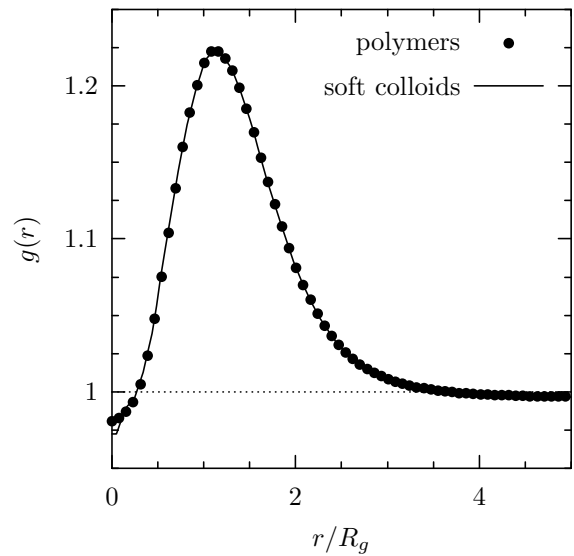}
\caption{\label{gr} Comparison of the pair distribution functions
obtained from simulations of a polymer system (symbols) and of the
corresponding soft colloid fluid in its homogeneous phase (solid
line). Here, $\beta=0.3$, $N=400$, $L=100$, $M=100^3$, and hence
$c=0.04$.}
\end{figure}

In Fig.~\ref{gr}, we compare the $g(r)$ of the metastable fluid phase
of the catastrophic soft colloid system to the $g(r)$ of the CM's of
the underlying polymer system at $c=0.04$ and $\beta=0.3$. The
agreement is excellent! This suggests that the HNC closure used in the
inversion procedure, a closure which has been shown to be very
accurate for bounded stable potentials
\cite{loubolhan00pre,lanlikwatlowjpcm00}, still works, in spite of the
fact that the potential it produces is catastrophic
\cite{roostiwas88jcp}. The situation is akin to that of the
Percus-Yevick solution for the structure of Baxter's sticky sphere
model \cite{bax68jcp}, which is useful for describing hard colloidal
systems with very short ranged attractive interactions in spite of the
fact that the underlying model system is actually catastrophic
\cite{ste91jsp}.  In both cases, the problem is most likely
circumvented by the approximate character of the chosen closure.

\subsection{Representability problems for effective potentials}

The considerations of the previous section, where the coarse-graining
procedure led to catastrophic potentials, are examples of the more
general problems of {\em transferability} and {\em representability}
of effective potentials used to describe complex systems, issues
discussed in more detail in a recent review \cite{lou02jpcm}.

Transferability problems occur when an effective potential derived at
one state point is not applicable at a different state point.  In
essence all the derived effective potentials in this paper suffer from
transferability problems, since they vary with density and with
temperature.  However, for high temperatures, the density dependence
is not that strong, so that the transferability problems are not as
important as they appear to be at lower temperatures.  For example,
the effective potentials in Fig.~\ref{dens3} vary much more rapidly
with density, suggesting that one must be very careful in using a
potential derived at one state point as an approximate one for another
state point.

Representability problems occur when effective potentials derived to
reproduce one physical property, do not easily describe another
physical property \cite{lou02jpcm}.  For example, one can calculate
the (unique) pair potential $v_g(r)$ that reproduces the homogeneous
structure (described by $g(r)$) of a system interacting via a
Hamiltonian with pair and triplet interaction terms (similarly to what
was done in this paper). If the usual equations for the internal
energy or the virial pressure, valid for normal pair potentials, are
applied to $v_g(r)$, then the latter thermodynamic quantities are not
correctly reproduced.  This was pointed out many years ago in the
context of simple fluids \cite{barhensmi69mp}. The same
representability problems were already found for the athermal polymer
case \cite{lou02jpcm}. They are expected to be more pronounced when
the density dependence is as prominent as in Fig.~\ref{dens3}.
Moreover, as pointed out in the previous section, even more vexing
representability problems occur, because the derived effective
potentials can lead to systems without a well-defined thermodynamic
limit.

The problems of transferability and representability are more
important for inhomogeneous systems. For example, at any interface
between two phases, it is not clear which potential should be used.
Similarly, for the temperature regime that leads to the catastrophic
potentials of the previous section, the apparent metastability of the
effective ``soft colloid'' system allows for the inversions to work
for a homogeneous system, but this breaks down if the simulations are
started with certain inhomogeneous initial conditions.  Using some
measure of the local density may be a better path to follow.  Taking
again the example of Fig.~\ref{dens3}, if one were to use a local
density dependence, the effective system could be stable against
collapse, as the potentials would become more repulsive for higher
local densities.  This would then more closely resemble the underlying
polymer system.  However, prescriptions for taking local density
dependence into account which are both accurate and tractable are not
yet well developed.

A final issue not yet resolved is the possible role of volume terms --
contributions to the free-energy which are independent of the
particular CM configuration.  Their effects on phase behavior can be
subtle (see e.g. \cite{gralow98pre,Liko01}), and they may appear in
certain coarse-graining schemes \cite{stisaktor02jcp}. For the case of
polymers in a good solvent, they were shown to be negligible
\cite{bollouhanmei01jcp,Liko01}, but that may no longer be the case
for poorer solvents.

\section{Conclusion}\label{conclusion}

We have extended previous work on a coarse-grained description of
polymer solutions in good solvent to increasingly poor solvent
conditions by adding nearest-neighbor attractions to the initial
lattice SAW model, and gradually lowering the temperature.  As in the
earlier work \cite{bollouhanmei01jcp,bollouhan01pre,bollou02mm}, an
effective pair potential $v(r)$ between the centers of mass of the
linear polymer coils was extracted from the Monte Carlo generated
CM-CM pair distribution function $g(r)$, using the HNC closure, which
is known to be very accurate in the absence of any hard core
repulsion.  In the infinite dilution limit, the effective pair
potential $v_2(r)$ is close to a Gaussian-shaped repulsion, of
amplitude $\beta v_2(0) \approx 2$, at high temperatures
(corresponding to the SAW limit), but this amplitude decreases as the
inverse temperature $\beta$ increases; an attractive tail develops and
at the lowest temperature investigated in our MC simulations, the
effective pair potential is entirely attractive, signalling a tendency
of the system to coalesce.

At these low temperatures, severe ergodicity problems arise in the MC
simulations, and, in the zero density limit, an overlapping histogram
method must be used to extract statistically significant results.  The
problem worsens with increasing polymer length $L$, so that our
simulations were mostly restricted to $L=100$.  

Increasing the polymer concentration at fixed temperature leads to a
``restabilization'' of the solution in the sense that the effective
pair potential exhibits an increasingly repulsive component as the
system is taken from the ultra-dilute to the semi-dilute regime.  Even
at the lowest temperature investigated ($\beta = 0.3$), the effective
pair potential reverts to an almost exclusively repulsive
Gaussian-like shape at $c=0.32$ (which, for these polymers, is close
to the melt regime).

The occurrence of strongly attractive and significantly
state-dependent effective pair potentials between polymer CM's raises
the question of the thermodynamic stability of systems of particles
interacting via such ``catastrophic'' potentials, and of the
appropriateness of the coarse-graining procedure to describe solutions
of interacting polymers, which are intrinsically stable, close to the
$\theta$ temperature.  The observation that the effective
coarse-grained pair potential $v(r)$ is capable of reproducing the
$g(r)$ (derived from monomer scale simulations) despite the
catastrophic nature of the pair potential, points to the possible
existence of metastable homogeneous states generated by these
potentials.  These metastable states appear to exhibit proper
thermodynamic extensivity properties, and may be stabilized against
ultimate coalescence by sufficiently high kinetic barriers.  If this
is indeed the case, the catastrophic effective pair potentials may
still provide a useful coarse-graining tool to describe homogeneous
states.  Strongly inhomogeneous states, generated by coalescence, lead
to widely different local densities, and hence would require the use
of effective pair potentials depending on the local density.  The fact
that the effective pair potential tends to become more repulsive at
higher density might provide the stabilizing mechanism for the
homogeneous state.

We plan to extend the present work to examine polymer CM density
profiles near interfaces, and to extract the osmotic equation of state
of polymer solutions as a function of concentration and temperature.

\acknowledgments
VK acknowledges support from the EPSRC under grant number GR/M88839
and AAL acknowledges support from the Isaac Newton Trust, Cambridge,
and the Royal Society. We thank Chris Addison for a careful reading of
the manuscript.

\appendix
\section{Overlapping histogram method for the computation of the 
effective potential between two chains}

We first introduce a reference system consisting of two polymer chains
similar to the original ones, except that the intermolecular
interaction does not include an attractive nearest-neighbor
contribution. The corresponding effective pair potentials
$v_2^{\text{ref}}(r)$ can be efficiently computed with the direct
method, since the intermolecular Boltzmann weights can only be either
$0$ or $1$ in this case.

The overlapping distribution method is then used to compute
$v_2(r)-v_2^{\text{ref}}(r)$, the free-energy difference between the
two original chains, with their CM's constrained to stay at a given
distance $r$, and the two reference chains under the same geometrical
constraint. To do so, we need to know for these systems the
distribution functions $\pi_{\text{orig}}(r,n_c)$ and
$\pi_{\text{ref}}(r,n_c)$, respectively, of the number $n_c$ of
intermolecular contacts \cite{note1}, which, according to the theory
of the overlapping distribution method, obey the relation
\begin{equation}\label{eqover}
\beta\,v_2(r)-\beta\,v_2^{\text{ref}}(r)=\ln
\frac{\pi_{\text{orig}}(r,n_c)}{\pi_{\text{ref}}(r,n_c)}+
\beta\varepsilon n_c,
\end{equation}
where $\varepsilon=-1$ denotes the nearest-neighbor pair attraction.

These histograms have been computed by sampling in canonical Monte
Carlo simulations, for both the original and the reference systems,
the ensemble of two chain configurations with CM's at distance $r$.
The elementary move we use is the following. One chain is chosen at
random and a pivot move is attempted on it. If no self-overlap,
leading to immediate rejection, occurs for this chain, a random
position with CM's at distance $r$ is chosen for the modified chain
around the other coil. If in this position the two chains overlap, the
move is rejected, else it is accepted or rejected according to the
standard Metropolis rules depending on the change in total energy.

\begin{figure*}
\includegraphics{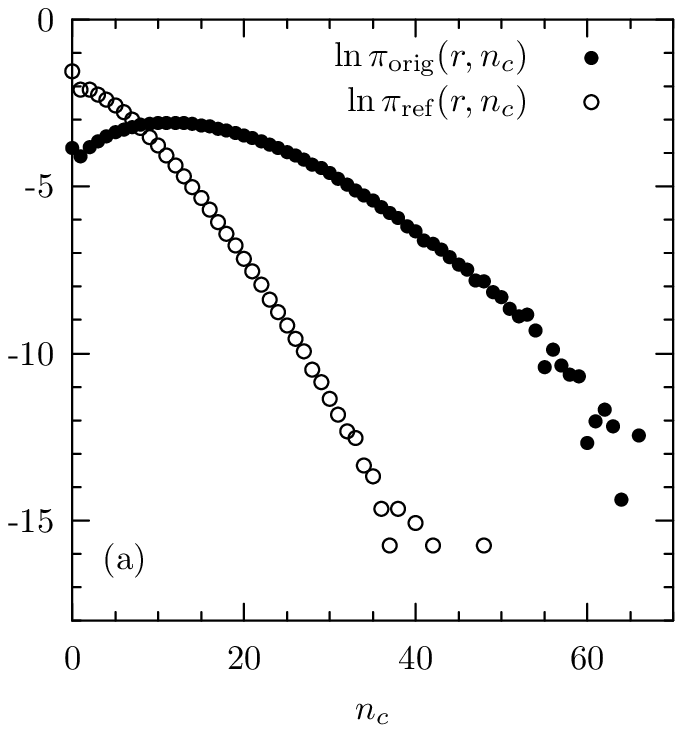}
\hspace{0.5cm}
\includegraphics{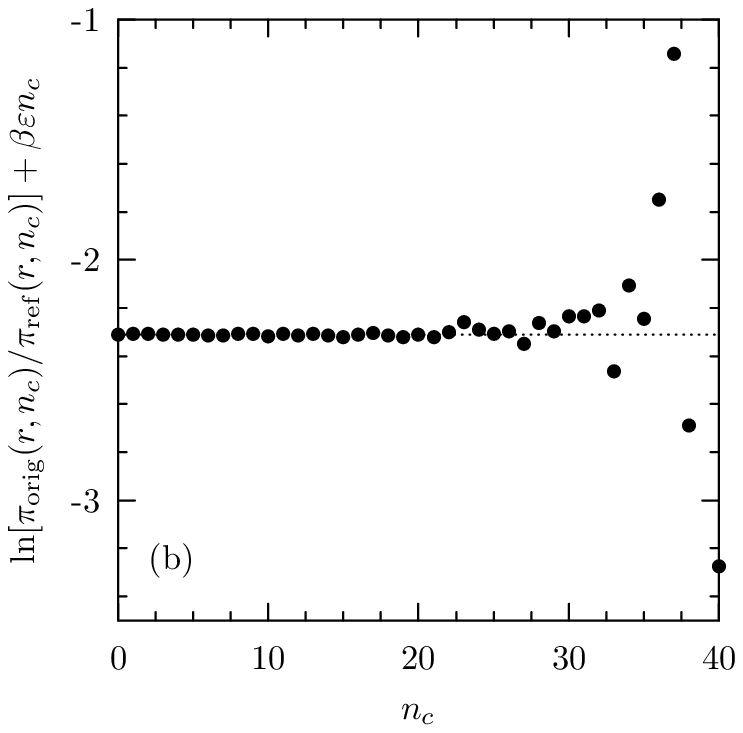}
\caption{\label{overlap2} Distribution functions for the number $n_c$
of intermolecular contacts between two simple cubic lattice chains of
length $L=100$ at $\beta=0.3$ and with CM's constrained to stay at
distance $r=4.2$. $\pi_{\text{orig}}(r,n_c)$ and
$\pi_{\text{ref}}(r,n_c)$ refer to the original and reference systems
respectively. (a) Log plot of $\pi_{\text{orig}}(r,n_c)$ and
$\pi_{\text{ref}}(r,n_c)$ as functions of $n_c$. (b) Determination of
the free energy difference $v_2(r)-v_2^{\text{ref}}(r)$ using
Eq.~\eqref{eqover}. As expected from the theory, no explicit
dependence of the combination
$\ln\left[\pi_{\text{orig}}(r,n_c)/\pi_{\text{ref}}(r,n_c)\right]
+\beta\varepsilon n_c$ on $n_c$ is found.}
\end{figure*}

Examples of the accumulated histograms, together with their
combination through formula \eqref{eqover}, are shown in
Fig.~\ref{overlap2}. Clearly, the explicit dependence of the left-hand
side of Eq.~\eqref{eqover} on $n_c$ is found to disappear as
prescribed by the theory, giving us a good indication of the
convergence of the method.

\begin{figure}
\includegraphics{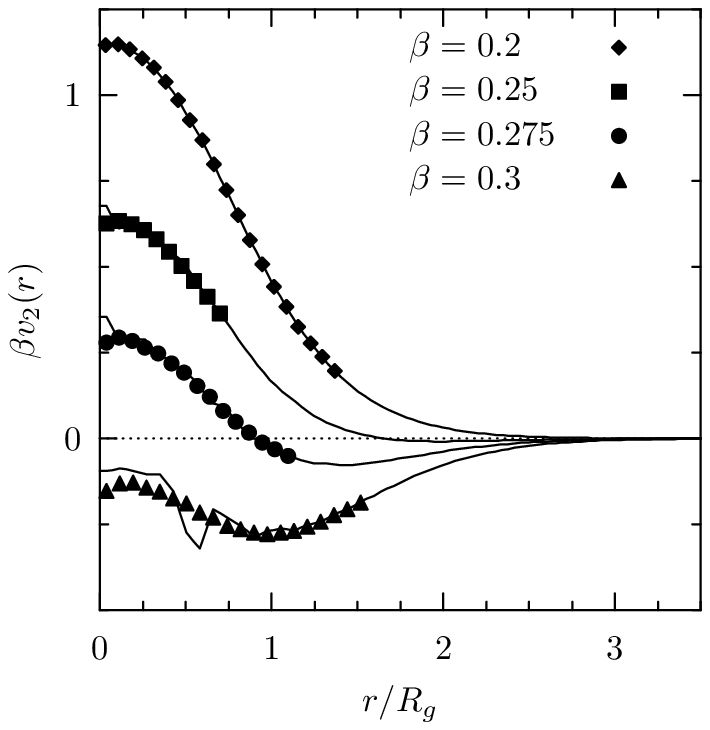}
\caption{\label{histo} Comparison of the zero density effective pair
potentials obtained with the overlapping histogram method (symbols)
and with the direct approach (solid lines) at various inverse
temperatures $\beta$ for polymers of length $L=100$ on the simple
cubic lattice.}
\end{figure}

The corresponding zero density pair potentials for polymer chains of
length $L=100$ are shown in Fig.~\ref{histo}, where they are compared
to the results obtained with the direct simulation method. $\beta=0.2$
is the largest value at which the direct method gives smooth and
reproducible results and thus has been used as a test case for the
overlapping histogram approach: as can be seen on the figure, the
curves obtained with both simulation methods are indistinguishable.
For larger $\beta$'s, the effective pair potentials computed with the
direct method display irregularities at small distances, especially
apparent for $\beta=0.3$, and originating in the sampling difficulties
mentioned in Sec.~\ref{zerodens}. This is clearly not the case of
those obtained with the histogram approach which are perfectly smooth
and meet the preceding curves for large enough distances at which
sampling problems disappear.

\clearpage

\begin{thebibliography}{99}
\bibitem{review} For a review, see K. Kremer, in \textit{Monte Carlo
and Molecular Dynamics of Condensed Matter Systems}, edited by
K. Binder and G. Ciccoti (SIF, Bologna, 1996).
\bibitem{degennes} P. G. de Gennes, \emph{Scaling Concepts in Polymer
Physics} (Cornell University Press, Ithaca NY, 1979).
\bibitem{flokri50jcp} P. J. Flory and W. R. Krigbaum, J. Chem. Phys.
\textbf{18}, 1086 (1950).
\bibitem{grokhakho82mcrc} A. Y. Grosberg, P. G. Khalatur, and A. R.
Khokhlov, Makromol. Chem. Rapid Commun. \textbf{3}, 709 (1982).
\bibitem{olalanpel80mm} O. F. Olaj, W. Lantschbauer, and K. H.
Pelinka, Macromolecules \textbf{13}, 299 (1980).
\bibitem{dauhal94mm} J. Dautenhahn and C. K. Hall, Macromolecules
\textbf{27}, 5399 (1994).
\bibitem{bollouhanmei01jcp} A. A. Louis, P. G. Bolhuis, J. P. Hansen,
and E. J. Meijer, Phys. Rev. Lett. \textbf{85}, 2522 (2000); P. G.
Bolhuis, A. A. Louis, J. P. Hansen, and E. J. Meijer, J. Chem. Phys.
\textbf{114}, 4296 (2001).
\bibitem{kruschbau89jpf} B. Kr\"{u}ger, L. Sch\"{a}fer, and A.
Baumg\"{a}rtner, J. Phys. France \textbf{50}, 319 (1989).
\bibitem{simpleliquids} see e.g. J. P. Hansen and I. R. McDonald,
\textit{Theory of Simple Liquids, 2nd Ed.} (Academic Press, London,
1986).
\bibitem{bollou02mm} P. G. Bolhuis and A. A. Louis, Macromolecules
\textbf{35}, 1860 (2002).
\bibitem{loubolmeihan02jcp1} A. A. Louis, P. G. Bolhuis, E. J. Meijer,
and J. P. Hansen, J. Chem. Phys. \textbf{116}, 10547 (2002).
\bibitem{loubolmeihan02jcp2} A. A. Louis, P. G. Bolhuis, E. J. Meijer,
and J. P. Hansen, J. Chem. Phys. \textbf{117}, 1893 (2002).
\bibitem{bollouhan} P. G.  Bolhuis, A. A. Louis, and J. P. Hansen,
Phys. Rev. Lett. \textbf{89}, 128302 (2002).
\bibitem{graheg95jcp} P. Grassberger and R. Hegger, J. Chem. Phys.
\textbf{102}, 6881 (1995).
\bibitem{ben76jcp} C. H. Bennett, J. Comput. Phys. \textbf{22}, 245
(1976).
\bibitem{understanding} D. Frenkel and B. Smit, \textit{Understanding
molecular simulations} (Academic Press, 1995).
\bibitem{lattice} A lattice model has been studied for computational
convenience. One can reasonably expect that qualitatively similar
results would be obtained for a continuous system, since the
properties investigated in this work have a characteristic length
scale of the order of $R_g$ and thus should not be too dependent on
the microscopic details of the model. A comparison of the results of
the present work with those of Dautenhahn and Hall \cite{dauhal94mm},
though rather limited, seems to confirm this point.
\bibitem{flory} P. J. Flory, \textit{Principles of Polymer Chemistry}
(Cornell University Press, Ithaca NY, 1953).
\bibitem{limadsok95jsp} B. Li, N. Madras, and A. Sokal, J. Stat. Phys.
\textbf{80}, 661 (1995).
\bibitem{bru84mm} W. Bruns, Macromolecules \textbf{17}, 2826 (1984).
\bibitem{meilim90jcp} H. Meirovitch and H. A. Lim, J. Chem. Phys.
\textbf{92}, 5144 (1990).
\bibitem{szlotopan92jcp} I. Szleifer, E. M. O'Toole, and
A. Z. Panagiotopoulos, J. Chem. Phys. \textbf{97}, 6802 (1992).
\bibitem{yandep00jcp} Q. Yan and J. J. de Pablo, J. Chem. Phys.
\textbf{113}, 5954 (2000).
\bibitem{gra97pre} P. Grassberger, Phys. Rev. E \textbf{56}, 3682
(1997).
\bibitem{fragra97jcp} H. Frauenkron and P. Grassberger, J. Chem. Phys.
\textbf{107}, 9599 (1997).
\bibitem{bax68jcp} R. J. Baxter, J. Chem. Phys.  \textbf{49}, 2270
(1968).
\bibitem{ste91jsp} G. Stell, J.  Stat.  Phys. \textbf{63}, 1203
(1991).
\bibitem{domjoy72jpc} C. Domb and G. S. Joyce, J. Phys. C \textbf{5},
959 (1972); C. Domb, J. Stat. Phys. \textbf{30}, 425 (1983).
\bibitem{duxquesti84jpa} P. M. Duxbury, S. L. A. de Queiroz, and R.
B. Stinchcombe, J. Phys. A \textbf{17}, 2113 (1984); P. M. Duxbury and
S. L. A. de Queiroz, J. Phys. A \textbf{18}, 661 (1985).
\bibitem{loubolhan00pre} A. A. Louis, P. G. Bolhuis, and J. P. Hansen,
Phys. Rev. E \textbf{62}, 7961 (2000).
\bibitem{ensemble} One can easily recognize that $\langle
W(|\mathbf{r}_{AB}|=r; \Gamma_A,\Gamma_B)\rangle$ is the ratio of the
partition function of two chains with their CM's constrained to stay
at distance $r$ over that of two chains infinitely far apart (which is
the square of the single chain partition function), and thus that
Eq.~\eqref{meanforce} is a direct implementation of the definition of
$v_2(r)$ as the free energy difference between these two systems. But
$\langle W(|\mathbf{r}_{AB}|=r; \Gamma_A,\Gamma_B)\rangle$ can also be
shown to be the zero density CM pair distribution function as
calculated in the grand-canonical ensemble, evidencing that $v_2(r)$
is also the potential of mean-force between the two coils.
\bibitem{madsok88jsp} N. Madras and A. D. Sokal, J. Stat. Phys.
\textbf{50}, 109 (1988).
\bibitem{wid63jcp} B. Widom, J. Chem. Phys. \textbf{39}, 2802 (1963). 
\bibitem{shigub82mp} K. S. Shing and K. E. Gubbins, Mol. Phys.
\textbf{46}, 1109 (1982); \textbf{49}, 1121 (1983).
\bibitem{grokuz92mm} Such a behavior has been seen in computer
simulations, see Ref.~\cite{olalanpel80mm}, and is predicted by the
mean-field theory as well, see for instance A. Y. Grosberg and D. V.
Kuznetsov, Macromolecules \textbf{25}, 1991 (1992).
\bibitem{semidilute} One difficulty with the $L=100$ chains is that
there is no well-defined semi-dilute regime, as discussed in more
detail in \cite{bollouhanmei01jcp}. When $(4/3)\pi R_g^3\rho\approx1$,
the monomer density is already $c \approx 0.1$, so that one rapidly
enters into the melt regime.
\bibitem{lanlikwatlowjpcm00} A. Lang, C. N. Likos, M. Watzlawek, and
H. L\"{o}wen, J. Phys.: Condens. Matter \textbf{12}, 5087 (2000);
C. N. Likos, A. Lang, M. Watzlawek, and H. L\"{o}wen, Phys. Rev. E
\textbf{63}, 031206 (2001).
\bibitem{ruelle} D. Ruelle, \emph{Statistical Mechanics: Rigorous
Results}, (W. A. Benjamin, Inc., London, 1969).
\bibitem{gallavotti} To avoid ambiguity, we use the word coalescence,
as in G. Gallavotti, \textit{Statistical Mechanics. A short treatise},
(Springer-Verlag, New York, 1999), to qualify the behavior of systems
interacting via catastrophic pair potentials, and reserve the word
collapse to refer to the coil-to-globule transition of an isolated
polymer chain.
\bibitem{hen74pla} R. L. Henderson, Phys. Lett. A \textbf{49}, 197
(1974); J. T. Chayes and L. Chayes, J. Stat. Phys. \textbf{36}, 471
(1984).
\bibitem{gralow98pre} R. van Roij and J. P. Hansen, Phys. Rev. Lett.
\textbf{79}, 3082 (1997); H. Graf and H. L\"owen, Phys. Rev. E
\textbf{57}, 5744 (1998); R. van Roij, M. Dijkstra, and J. P. Hansen,
Phys. Rev. E \textbf{59}, 2010 (1999); M. Dijkstra, R. van Roij, and
R. Evans, Phys.  Rev. E \textbf{59}, 5744 (1999).
\bibitem{Liko01} C. N. Likos, Phys. Rep. \textbf{348}, 267 (2001).
\bibitem{rea86pma} L. Reatto, Phil. Mag. A \textbf{58}, 37 (1986); L.
Reatto, D. Levesque, and J. J. Weis, Phys. Rev. A. \textbf{33}, 3451
(1986).
\bibitem{zerhan86jcp} G. Zerah and J. P. Hansen, J. Chem. Phys.
\textbf{84}, 2336 (1986).
\bibitem{milpaubin93jcp} A. Milchev, W. Paul, and K. Binder,
J. Chem. Phys. \textbf{99}, 4786 (1993); O. F. Olaj, T. Petrik, and
G. Zifferer, Macromol. Theory Simul. \textbf{6}, 1277 (1997);
J. Chem. Phys. \textbf{107}, 10214 (1997).
\bibitem{bollouhan01pre} P. G. Bolhuis, A. A. Louis, and J. P. Hansen,
Phys. Rev. E \textbf{64}, 021801 (2001).
\bibitem{lou02jpcm} A. A. Louis, J. Phys.: Condens. Matter
\textbf{14}, 9187 (2002).
\bibitem{more} The use of the approximate criterion \eqref{eqA3} might
have a deeper physical meaning.  First, it can be seen as a
specialization of the general stability criterion to a restricted
configuration space from which inhomogeneous, non fluid-like
configurations are excluded, motivated by the fact that the original
polymer system is always fluid and homogeneous in the conditions of
the present study. Similar constructions are often invoked in various
fields of liquid state theory, for instance to study systems
metastable with respect to crystallization like supercooled liquids or
colloids near their putative liquid-gas transition.  Second, $I_2$ and
$B_2$ are similar simple functionals of $\beta v_2$. Now, $B_2$ is
used \emph{universally}, i.e. independently of the details of the
molecular structure and of the microscopic interactions, for the
characterization of the $\theta$ point. The use of $I_2$ could share
this universal character. In this respect, it is interesting to note
that in the Domb-Joyce model the coalescence catastrophe of a many
polymer system, accompanying the appearance of self-trapping behavior
of individual chains, occurs exactly when $I_2$ vanishes.
\bibitem{roostiwas88jcp} These considerations contrast with the
approach of L. J. Root, F. H. Stillinger, and G. E.  Washington,
J. Chem. Phys. \textbf{88}, 7791 (1988), who, from the strictly
opposite point of view, proposed to use the limited ability of a
closure relation to detect the coalescence catastrophe embedded in a
given potential as a criterion to evaluate the quality of the
corresponding integral equation theory.
\bibitem{barhensmi69mp} J. A. Barker, D. Henderson, and W. R. Smith,
Mol. Phys. \textbf{17}, 579 (1969); G. Casanova, R. J. Dulla, D. A.
Jonah, J. S. Rowlinson, and G. Saville, Mol. Phys. \textbf{18}, 589
(1970); J. S. Rowlinson, Mol. Phys. \textbf{52}, 567 (1984); M. A.
van der Hoef and P. A. Madden, J. Chem. Phys. \textbf{111}, 1520
(1999).
\bibitem{stisaktor02jcp} F. H. Stillinger, H. Sakai, and S. Torquato,
J. Chem. Phys. \textbf{117}, 288 (2002).
\bibitem{note1} In the general theory of the overlapping distribution
method, distribution functions for the energy differences between the
original and reference systems are needed, but here, with the system
at hand and for our choice of the reference, the identity between
these energy differences and the number $n_c$ of intermolecular
contacts is evident.

\end{thebibliography}
\end{document}